\documentclass{iaus}
\newcommand\msolar{M$_\odot$}
\newcommand\apj{\textit{ApJ}}                
\newcommand\apjl{\textit{ApJ (Letters)}}                
\newcommand\aap{\textit{A\&A}}                
\newcommand\mnras{\textit{MNRAS}}              
\usepackage{graphicx}
\title[Bars in Cuspy Dark Halos]{Bars in Cuspy Dark Halos}

\author[Dubinski, Berentzen, \& Shlosman]{
John Dubinski$^1$,
Ingo Berentzen$^2$ and,
Isaac Shlosman$^{3,4}$
}

\affiliation{$^1$Department of Astronomy and Astrophysics,
University of Toronto, \\ 
50 St. George Street, Toronto, ON M5S 3H4, Canada \\ email: 
{\tt dubinski@astro.utoronto.ca}\\[\affilskip]
$^2$Astronomisches Rechen-Institut, Zentrum f\"ur Astronomie,
      Universit\"at Heidelberg
M\"onchhofstr. 12-14 69120,
Heidelberg, Germany \\ email: {\tt iberent@ari.uni-heidelberg.de}\\[\affilskip]
$^3$JILA, University of Colorado, Boulder, CO 80309-0440, USA; and \\
$^4$Department of Physics and Astronomy, University of
Kentucky, \\ Lexington, KY 40506-0055, USA\\ email: {\tt shlosman@pa.uky.edu}}

\pubyear{2008}
\volume{254}  
\pagerange{1-6}
\jname{The Galaxy Disk in Cosmological Context}
\editors{J.~Andersen,J.~Bland-Hawthorn, B.~Nordst\"om}
\begin{document}

\maketitle

\begin{abstract}
We examine the bar instability in models with an exponential disk and a cuspy NFW-like dark matter (DM)
halo inspired by cosmological simulations.  Bar evolution is studied as a function of
numerical resolution in a sequence of models spanning $10^{4-8}$ DM particles - including a multi-mass
model with an effective resolution of $10^{10}$.  The goal is to find convergence 
in dynamical behaviour.  
We characterize the bar growth, the buckling instability, pattern speed decay through resonant transfer
of angular momentum, and possible destruction of the DM halo cusp.
Overall, most characteristics converge in behaviour for halos containing more than $10^7$ particles
in detail.  Notably, the formation of the bar does not destroy the density cusp in this case.
These higher resolution simulations clearly illustrate the importance of discrete resonances in
transporting angular momentum from the bar to the halo.
\end{abstract}

\firstsection 

\vspace{0.5cm}
\noindent {\bf 1.~Introduction -- }
The bar instability in a cold gravitating disk plays a major role in a spiral galaxy's
dynamical evolution.  
At least 2/3 of spiral galaxies host bars (\cite[Knappen \etal~2000]{kna00})
and the fraction has not evolved significantly since $z\sim
1$ (\cite[Jogee \etal~2004]{jo04}; \cite[Sheth \etal~2008]{she08}).  
As models of galaxy formation become more sophisticated and
reveal complex dynamical behaviour, 
it is important to understand the details of different
physical processes that shape their morphology as well as to verify that numerical
resolution is in fact adequate to follow their evolution.  The bar-halo interaction is the
driving mechanism in disk galaxy evolution.  As a bar churns through the DM halo
with a pattern speed $\Omega_b$ resonant interactions with halo orbits -- a form of
dynamical friction -- transfer
angular momentum from the bar to the halo and cause it to spin down (\cite[Tremaine \& Weinberg
1984]{tre84}).  This process was first pointed out by \cite[Lynden-Bell \& Kalnajs (1972)]{lyn72}
and has been studied in models with idealized rigid bars 
(\cite[Weinberg 1985]{wei85}; \cite[Hernquist \& Weinberg 1992]{her92}; \cite[Weinberg \& Katz
2002]{wei02}; 
\cite[Weinberg \& Katz 2007]{wei07a}) as well as in models 
in which a stellar bar forms 
self-consistently in an unstable disk 
(e.g., \cite[Sellwood 1980]{sel80}; \cite[Debattista \& Sellwood 1998]{deb98};
\cite[O'Neill \& Dubinski 2003]{one03}; 
\cite[Holley-Bockelmann \etal~2005]{hol05}; 
\cite[Martinez-Valpuesta \etal~2006]{mar06}).   
There has been some concern
that the process is too efficient, leading to bars that are much smaller than their corotation
radii and so discrepant with observed bar galaxies (\cite[Debattista \& Sellwood
2000]{deb00}).  More recent studies with greater resolution
suggest that the bars tend to lengthen moving out to their co-rotation radii as they slow down
and so perhaps they are not inconsistent with reality 
(\cite[O'Neill \& Dubinski 2003]{one03}; 
\cite[Martinez-Valpuesta \etal~2006]{mar06}).
\cite[Weinberg \& Katz (2002)]{wei02} hold a cautious view that lower resolution simulations
can lead to spurious results because of the diffusive nature of  noise that may move orbits 
into and out of resonances artificially
while insufficient particle numbers may also underpopulate the resonant regions of phase space.  While
most current work to date has used $\sim 10^6$ particles, they claim that as many as $~10^8$
DM particles (or more) may be needed to simulate the resonant process with $N$-body methods.

In this study, we attempt to clear up the inconsistencies of current work and address the
problem of numerical resolution hoping to converge to the correct physical behaviour.
We present a series of bar-unstable disk$+$halo $N$-body models with increasing resolution
spanning a range $N_h=10^4-10^8$ DM particles with $N_d=1.8\times 10^{3-7}$ disk particles.
One further simulation uses a multi-mass method that increases the halo particle number density
by $200\times$ in the halo centre so giving an effective $N_h\sim 10^{10}$.  The mass model is
constructed from a 3-integral distribution function (\cite[Widrow \& Dubinski 2005]{wid05})
describing an exponential disk embedded within a tidally truncated NFW halo and is similar
to the model studied by
\cite[Martinez-Valpuesta \etal~(2006)]{mar06}.   
Natural units for the simulations are $D=10$ kpc, $M=10^{11}$ \msolar,
$V=208$ km/s, and $T=47.2$ Myr.  We discuss results both in simulation and physical units
throughout and clarify when necessary.  
(For further details on the models and simulations
see \cite[Dubinski, Berentzen \& Shlosman 2008]{dub08}).
Animations of the simulations are available for viewing at the URL:
\verb www.cita.utoronto.ca/~dubinski/IAU254 ~along with higher-resolution figures.

Animation~1 shows the evolution of the disks in six models with increasing resolution in
face-on and edge-on views.  The lowest resolution simulations with $N_h \le 10^5$
are clearly deficient and either lose the bar or suffer from heating affects.  At higher
resolution, the behaviour is similar exhibiting the buckling instability and relaxation to a
bar in quasi-equilibrium that gradually lengthens and slows down.  Animations 2 \& 3 show the
multi-mass model with $N_h=10^8$ in an inertial and corotating frame and illustrate how the bar
grows from noise from the inside out saturating as a thin bar on reaching the corotation radius
and then evolving into a fatter bar with a peanut-shaped bulge after the buckling instability.

\begin{figure}[t]
\begin{center}
\includegraphics[width=4.5in]{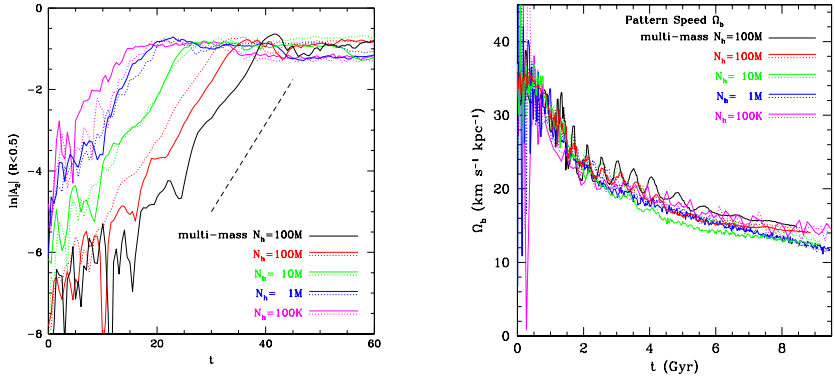} 
\caption{
Initial growth of the $m=2$ Fourier component $|A_2|$ for stars with $R<0.5$
for two model sequences using
$N_d=18K,180K,1.8M,18M$ with $N_h=100K,1M,10M,100M$ respectively.
The $\ln|A_2|$ grows approximately linearly with time independent
of the choice of $N_d$ and $N_h$ showing the
exponential growth of the bar mode.  The dashed line shows an exponential
timescale that is approximately $\tau=8$ (370~Myr). Since the bar grows
from the Poisson noise within the disk then we expect the
noise amplitude to be proportional to $N^{1/2}$ so larger simulations will saturate at later
times.
\label{fig-lseq}
}
\end{center}
\end{figure}

\vspace{0.5cm}
\noindent {\bf 2.~Bar Growth and Pattern Speed Evolution --}
We study bar growth using the normalized Fourier amplitude $|A_2|$ of the $m=2$ 
disturbance in the plane
of the disk within a fixed radius $R<0.5$ (5 kpc).  Figure~\ref{fig-lseq} shown the exponential
growth of $|A_2|$ before saturation.  Higher resolution simulations reach saturation at later times.
This time delay is the result of the lower amplitude of the Poisson fluctuations that 
seed the bar.  Since the instability grows from these fluctuations if takes longer for them to
saturate if the initial amplitude is lower.  We estimate the time delays from the peak in $|A_2|$
and synchronize the simulations for comparison of various evolutionary characteristics.  

\begin{figure}[t]
\begin{center}
\caption{
Evolution of the pattern speed $\Omega_b$ for two model sequences at different resolution.
The curves have been shifted
in time so that the bar growth evolution is coincident with the 1M particle  model.
\label{fig-omega}
}
\end{center}
\end{figure}

We can also use $A_2$ to measure the phase angle of the $m=2$ mode and so estimate the pattern speed
by taking the difference between subsequent snapshots.  Figure~\ref{fig-omega} shows the pattern
speed evolution as a function of numerical resolution.  There is some scatter in behaviour that
can be accounted for from the expected variance introduced by different initial conditions but
overall the agreement is good.  The highest resolution simulations show a modulation of the
pattern speed at a frequency close to $\Omega_b$ itself.  This probably indicates an interaction
between the bar and the gravitational wake in the halo that only shows up with sufficient
numerical resolution.

\begin{figure}[tb]
\begin{center}
\includegraphics[width=3.4in]{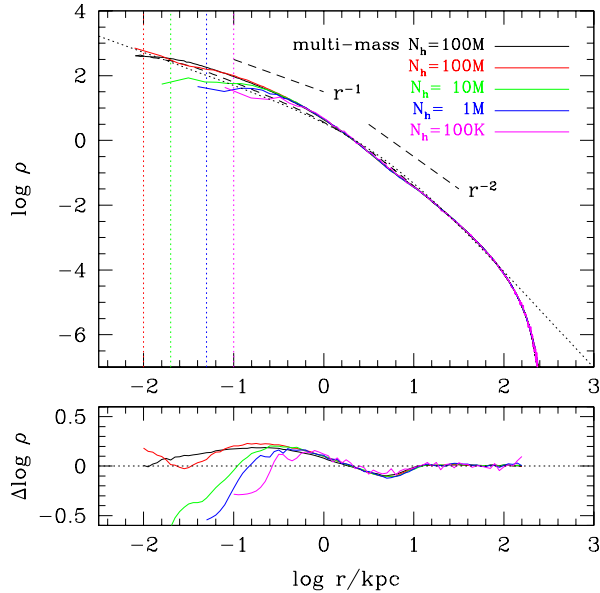} 
\caption{
A comparison of density profiles at $t=7.08$~Gyr for different halo
particle numbers $N_h$.  We also show the initial density profile (dashed
line) and the best fit NFW model curve (dotted line) to the initial
profile over the range $0<r<100$~kpc.
The NFW parameters for the fit are $r_s=4.3$ kpc, $v_{max}=160$~km~s$^{-1}$,
where
$v_{max}$ is the maximum circular velocity at $r=2.16 r_s=9.3$~kpc.
Note that this halo is more concentrated than the typical
galactic dark matter halos in cosmological simulations to model the
contraction expected during dissipative galaxy formation.
The dotted vertical lines show the softening length $\epsilon$ used at
different resolutions.
\label{fig-den}
}
\end{center}
\end{figure}

\pagebreak
\noindent {\bf 3.~Halo Density Profile --}
We also measured the evolution of the halo density profile over the course of the simulation.
Previous work has shown both preservation and destruction of the density cusp
so we focus on the processes in the central regions.  
Only studies using the self-consistent field $N$-body method (SCF) lead to a core
(\cite[Holley-Bockelmann \etal~2005]{hol05}; \cite[Weinberg \& Katz 2007b]{wei07b}) and there has
been some concern about numerical instabilities that arise in those methods 
(\cite[Selwood 2003]{sel03}).
Figure~\ref{fig-den} exhibits the final density
profile along with the difference from the initial profile as a function of halo particle number.
The profiles agree well to within the limit of their gravitational softening radius and show
convergent behaviour.  The cusp is not destroyed in this case but rather the central density
increases modestly, by a factor of 2, as a result of the bar evolution, following its
buckling, which leads to an increase in the central stellar density (\cite[Sellwood 2003]{sel03}).  

The bar that forms in this simulation has a similar mass ratio $M_b/M_{halo}\approx 0.6$
but is much thicker than the fiducial rigid bar simulation in Weinberg \etal~(2007b)
that destroys the cusp.   Their study also included thick bars which did not destroy the cusps
and the bar that forms here overlaps with those in their study.  We conclude that there is no
direct contradiction with the most recent results of Weinberg \etal~(2007b) but would argue that 
the thicker bar models are probably more relevant to real galaxies.  Thin bars do not persist
for long before responding to the buckling instability and so the rigid bar approximation is not
applicable over a Hubble time and probably not relevant to most galaxies
(\cite[Martinez-Valpuesta \& Shlosman 2004]{mar04}).  

\vspace{0.5cm}
\noindent {\bf 4.~Orbital Resonances --}
The bar slowdown is the result of dynamical friction that leads to angular momentum transport to the DM halo.  
The process is due to resonant interactions between
the rotating bar's pattern speed and the halo particles' azimuthal and radial orbital frequencies.
When $l_1\Omega_r + l_2\Omega_\phi = m\Omega_b$ then orbital resonances occur and halo particles
torque or are torqued by the bar leading to a change in angular momentum.
We can estimate the orbital frequencies at a specific time by freezing the potential and
integrating the orbits of particles in this potential rotating with the pattern speed at that
time (e.g., \cite[Athanassoula 2002]{ath02}; \cite[Martinez-Valpuesta \etal~2006]{mar06}).  
Spectral analysis can then be 
applied to the orbital time series to determine the
fundamental orbital frequencies $\Omega_r$ and $\Omega_\phi$ (\cite[Binney \& Spergel 1982]{bin82}). 
(Note we label these frequencies
with the usual epicyclic variables $\kappa\equiv\Omega_r$ and $\Omega \equiv \Omega_\phi$ in our
figures below.)  The dimensionless frequency $\eta=(\Omega - \Omega_b)/\kappa$ is a useful way to
characterize resonances since the values $\eta=1/2,0,-1/2$ correspond to the inner Lindblad,
corotation, and outer Lindblad resonances for $m=2$.  Further negative half-integer 
values correspond to higher order resonances that can also absorb the angular momentum.

\begin{figure}[t]
\begin{center}
\includegraphics[width=4.0in]{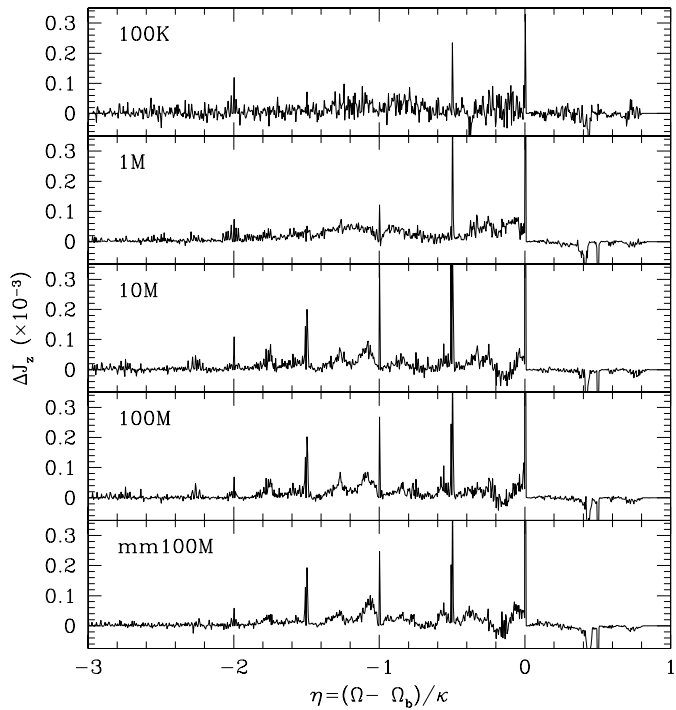} 
\caption{
The distribution in the 
change in the $z$ angular momentum $\Delta J_z$ between $t=100$ (4.7 Gyr) and $t=150$ (7
Gyr) plotted as a function of the dimensionless frequency $\eta=(\Omega-\Omega_b)/\kappa$
measured at $t=100$.
The spikes at $\eta=0.5,0.0,-0.5$ correspond to the ILR, COR,  and OLR
respectively while other spikes refer to higher order resonances.
This plot shows how halo particles in resonant orbits are the main sink of angular momentum.  
The detailed distributions are converging for $N_h>10^7$ particles while lower  resolution
simulation miss some of the higher order resonances.
\label{fig-dj1}
}
\end{center}
\end{figure}

Figure~\ref{fig-dj1} shows the change in the $z$ component of the angular momentum $J_z$
for particles binned as a
function $\eta$ between $t=100$ and $t=150$ at different numerical resolution.  
The spikes at the half-integer values reveal the resonances.  
Most angular momentum is transferred through the corotation resonance though
it is also appears to be transferred at other frequencies, but more randomly  ---
this difference is probably the result of particles in resonance before or after $t=100$.   The detailed
behaviour of the distributions seem to converge for $N_h \ge 10^7$.  At lower resolutions, the
resonant spikes have a smaller amplitude and higher order resonant interactions are missing.  

\begin{figure}[t]
\begin{center}
\includegraphics[height=2.5in]{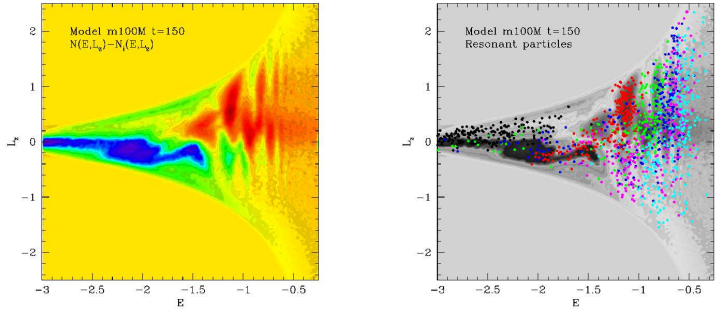} 
\caption{
Change in particle number density in $(E,L_z)$ space between $t=0$ and $t=150$ (7.0 Gyr) for the
$N_h=10^8$ single mass model.  The resonant regions show up clearly as peaks (red regions) in
phase space in the left panel.  In the right panel, we overplot the $(E,L_z)$ coordinates of
a random subset of particles
located at discrete resonances at $t=150$ within $\delta\eta=\pm 0.05$
(black-ILR-$\eta=0.5$, red-COR-$\eta=0.0$, green-OLR-$\eta=-0.5$, 
blue-$\eta=-1.0$, magenta-$\eta=-1.5$, and cyan-$\eta=-2.0$
\label{fig-phase}
}
\end{center}
\end{figure}

\pagebreak
Finally, we examine the change in halo phase space density by computing the particle number
density in $(E,L_z)$ space and computing the difference between $t=0$ and $t=150$ in a similar way
to \cite[Holley-Bockelmann \etal~(2005)]{hol05}.  In this way,
we clearly see the resonant regions visible as discrete islands of particle overdensity in
$(E,L_z)$ space (Fig.~\ref{fig-phase}).  We can also overplot the values of $(E,L_z)$ for the
particles found in the resonant spikes in the analysis to see where they lie in phase space.
The right panel of Figure~\ref{fig-phase} clearly shows that these islands are directly related
to the discrete resonances extracted in our spectral analysis.  Animation 4 describes the time
evolution of the differential number density in phase space and reveals how the resonances move
through a large fraction of the halo mass.  By counting particles in resonant peaks at different
times we estimate that roughly 30\% of the halo particles are affected.

We conclude that the resonances are broader than thought and so simulations with more than 1M
halo particles do a reasonable job of tracking bar evolution.  However, a look at the
distribution of orbital frequencies reveals that higher order resonances are missed at lower
resolution with less than 10M particles.  This effect could account for the
different rate of angular momentum loss at higher resolution.  Future studies should examine the
bar instability self-consistently using the same initial conditions with different $N$-body methods to resolve
current inconsistent results on the cusp/core evolution of DM halos.

%


\begin{thebibliography}{24}
\expandafter\ifx\csname natexlab\endcsname\relax\def\natexlab#1{#1}\fi


\bibitem[{{Athanassoula}(2002)}]{ath02}
{Athanassoula}, E. 2002, \apjl, 569, L83

\bibitem[{{Binney} \& {Spergel}(1982)}]{bin82}
{Binney}, J., \& {Spergel}, D. 1982, \apj, 252, 308

\bibitem[{{Debattista} \& {Sellwood}(1998)}]{deb98}
{Debattista}, V.~P., \& {Sellwood}, J.~A. 1998, \apjl, 493, L5+

\bibitem[{{Debattista} \& {Sellwood}(2000)}]{deb00}
---. 2000, \apj, 543, 704

\bibitem[{{Dubinski}(1996)}]{dub96}
{Dubinski}, J. 1996, New Astronomy, 1, 133

\bibitem[{{Dubinski} {et~al.}(2008){Dubinski}, {Berentzen}, \&
  {Shlosman}}]{dub08}
{Dubinski}, J., {Berentzen}, I., \& {Shlosman}, I. 2008, in prep.

\bibitem[{{Hernquist} \& {Weinberg}(1992)}]{her92}
{Hernquist}, L., \& {Weinberg}, M.~D. 1992, \apj, 400, 80

\bibitem[{{Holley-Bockelmann} {et~al.}(2005){Holley-Bockelmann}, {Weinberg}, \&
  {Katz}}]{hol05}
{Holley-Bockelmann}, K., {Weinberg}, M., \& {Katz}, N. 2005, \mnras, 363, 991

\bibitem[{{Jogee}(2004)}]{jog04etal}
{Jogee}, S., \etal~2004, \apjl, 615, L105

\bibitem[{{Knapen} {et~al.}(2000){Knapen}, {Shlosman}, \& {Peletier}}]{kna00}
{Knapen}, J.~H., {Shlosman}, I., \& {Peletier}, R.~F. 2000, \apj, 529, 93

\bibitem[{{Lynden-Bell} \& {Kalnajs}(1972)}]{lyn72}
{Lynden-Bell}, D., \& {Kalnajs}, A.~J. 1972, \mnras, 157, 1

\bibitem[{{Martinez-Valpuesta} \& {Shlosman}(2004)}]{mar04}
{Martinez-Valpuesta}, I., \& {Shlosman}, I. 2004, \apjl, 613, L29

\bibitem[{{Martinez-Valpuesta} {et~al.}(2006){Martinez-Valpuesta}, {Shlosman},
  \& {Heller}}]{mar06}
{Martinez-Valpuesta}, I., {Shlosman}, I., \& {Heller}, C. 2006, \apj, 637, 214

\bibitem[{{O'Neill} \& {Dubinski}(2003)}]{one03}
{O'Neill}, J.~K., \& {Dubinski}, J. 2003, \mnras, 346, 251

\bibitem[{{Sellwood}(1980)}]{sel80}
{Sellwood}, J.~A. 1980, \aap, 89, 296

\bibitem[{{Sellwood}(2003)}]{sel03}
---. 2003, \apj, 587, 638

\bibitem[{{Sheth}(2008)}]{she08etal}
{Sheth}, K. \etal~2008, \apj, 675, 1141

\bibitem[{{Tremaine} \& {Weinberg}(1984)}]{tre84}
{Tremaine}, S., \& {Weinberg}, M.~D. 1984, \mnras, 209, 729

\bibitem[{{Weinberg}(1985)}]{wei85}
{Weinberg}, M.~D. 1985, \mnras, 213, 451

\bibitem[{{Weinberg} \& {Katz}(2002)}]{wei02}
{Weinberg}, M.~D., \& {Katz}, N. 2002, \apj, 580, 627

\bibitem[{{Weinberg} \& {Katz}(2007{\natexlab{a}})}]{wei07a}
---. 2007{\natexlab{a}}, \mnras, 375, 425

\bibitem[{{Weinberg} \& {Katz}(2007{\natexlab{b}})}]{wei07b}
---. 2007{\natexlab{b}}, \mnras, 375, 460

\bibitem[{{Widrow} \& {Dubinski}(2005)}]{wid05}
{Widrow}, L.~M., \& {Dubinski}, J. 2005, \apj, 631, 838

\end{thebibliography}

\end{document}